# Environmental high resolution electron microscopy and applications to chemical science

## E.D. Boyes, P.L. Gai

*Central Research and Development, Science and Engineering Laboratories, DuPont, Experimental Station, Wilmington, DE 19880-0356, USA*

## Abstract

An environmental cell high resolution electron microscope (EHREM) has been developed for in situ studies of dynamic chemical reactions on the atomic scale. It allows access to metastable intermediate phases of catalysts and to sequences of reversible microstructural and chemical development associated with the activation, deactivation and poisoning of a catalyst. Materials transported through air can be restored or recreated and samples damaged, e.g. by dehydration, by the usual vacuum environment in a conventional electron microscope can be preserved. A Philips C M30 HRTEM/STEM system has been extensively modified in our laboratory to add facilities for in situ gas-solid reaction studies in controlled atmospheres of gas or vapor at pressures of 0-50 mbar, instead of the regular TEM high vacuum environment. The integrated new environmental cell capability is combined with the original 0.23 nm TEM resolution, STEM imaging (bright field/annular dark field) and chemical and crystallographic microanalyses. Regular sample holders are used and include hot stages to > 1000°C. Examples of applications include direct studies of dynamic reactions with supported metal particle catalysts, the generation of defects and structural changes in practical complex oxide catalyst systems under operating conditions and carbon microstructures.

## 1. Introduction

Direct observation of microstructural evolution under dynamic reaction conditions is a powerful scientific tool in materials science. In situ electron microscopy (EM) under controlled reaction conditions provides dynamic information on processes which cannot be obtained directly by other techniques. We are particularly interested in the properties of heterogeneous catalysts used in or proposed for commercial reactions, which may include environmental control as well as products, and to explore structure-property relationships associated with them [1]. A number of notable in situ experiments have relied upon modifications to the standard TEM operations. The main electron optical functions of the EM, especially the electron gun, depend on a high vacuum environment. The typical $10^{-6}$-$10^{-7}$ mbar TEM environment is mildly reducing. With the environmental cell (ECELL), controlled chemically reducing atmospheres (e.g. H2, hydrocarbons, CO) and oxidizing atmospheres (e.g. $O_2$), or solvent-rich (He/water vapor) are maintained in the EM. A wide range of gases and vapors can be used.

Some of the earlier in situ experiments carried out in high vacuum in the TEM have produced important results. The value of using a hot stage to study in situ crystallization with lattice imaging in a high vacuum environment of a TEM has been demonstrated [2]. Incorporation of a genuine UHV range suitable for surface science studies was started by Pashley [3] and has been developed into a powerful scientific tool by Yagi and coworkers [4]. Elegant experiments have been performed and the results of these studies are relevant to the wider surface science community. Venables et al. [5] have performed surface science investigations with a series of specialized SEM instruments. In the TEM the use of UHV methods has been combined with the highest resolution of which the instrument is capable.

However, there are important applications in materials science in which the role of environment on a sample is critical. Some of the earlier applications, summarized by Butler and Hale [6], were to provide a water saturated environment to prevent degradation of a solvated sample. Hydration of cement was carried out in a novel sample holder in an otherwise unmodified microscope [7]. The sample was sandwiched between thin windows to contain the saturated water vapor environment necessary both for preservation and reaction of the sample. Similar approaches have been employed to contain hydrated biological samples [8]. These and other workers made use of the high penetrating power and relatively large space in the high voltage electron microscopes (HVEM) introduced around 1970 [9], but rarely available today.

Electron-transparent windows have been used to contain gases, solvent vapors [7, 8] and UHV environments in microscopes of various voltages, but there are problems in reliably sustaining a large pressure difference across a window which is thin enough to permit electron penetration. They are acceptable at moderate resolution, but the additional diffuse scattering in the windows may further obscure the already limited image contrast. Superimposition of an additional diffraction pattern can be a problem and window-cells are generally not compatible with heating systems. In principle, the whole of a column of a microscope could be made UHV compatible, although it rarely is. Where the controlled atmosphere is comprised of gas or vapor it is important to limit the pathlength for additional electron scattering in the gas, and hence the length of the ECELL controlled environment zone, to a length of a few millimeters. This is also the configuration typically used for (non-vacuum generators whole column) UHV systems.

The complications and potential for catastrophic failure of windows can be avoided by substituting small apertures above and below the sample to restrict the diffusion of gas molecules, but to allow the penetration of the electron beam. Unlike windows, aperture systems are robust and they can easily be made compatible with sample heating or cooling; sometimes even using regular specimen holders for the purpose. However, there are two problems to consider. Firstly, the pressure differential which can be sustained across a single aperture (of a useful size and sealed around the perimeter) is limited to a factor of 100 x [6]. In practice, a series of apertures, with differential pumping systems connected between them, are needed above and below the sample (Fig. 1) to get from a minimum useful vapor pressure of e.g. 0.001-1 mbar around the sample, back to a maximum of $10^{-6}$ mbar in the electron gun. Typically, pairs of apertures are added above and below the sample with pumping lines attached between them [10], Most previous ECELL systems have used older [10], or larger [7-11] electron microscopes, some with inherent performance limitations and alignment problems of the basic instruments which were exacerbated by the design of ECELL systems and the attachment of the attendant pumping lines. Furthermore, many of these systems were made interchangeable, necessitating the frequent rebuilding of the microscope to effect the change-over in functionality. To maintain reliability, minimally invasive ECELL systems were designed which could "simply" be inserted between the polepieces of the objective lens. Such systems were, however, a major step forward in scientific

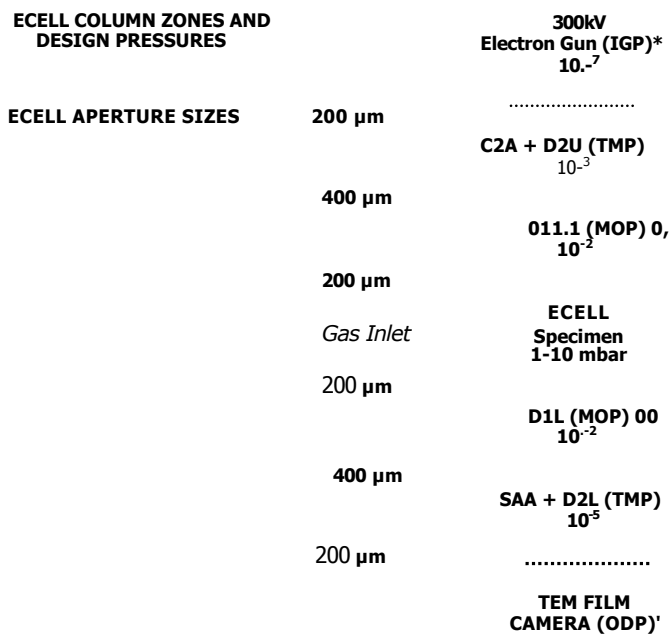

Fig. I. Schematic of ECELL HREM (CM30) column zones and design pressures. The notations are as follows: IGP = ion getter pump*; TMP = turbo molecular pump**; MDP = molecular drag pump*; ODP = oil diffusion pump + mechanical back pump; (* = original CM30 equipment; # = backed by diaphragm pump; # # = backed by New pump systems are oil-free gate valve added to isolate main (gun) **IGP** from cell atmosphere. The four primary ECELL apertures are mounted in bushes inside the bores of the objective lens polepieces.

capability and excellent work was carried out with them [1, 6-8, 24].

## 2. Experimental

Recently, a new approach has been taken to organize and design instruments which are dedicated to ECELL operations and remain in this configuration for long periods of time. In our case the modifications of the instrument are considered permanent. Firstly, it is based on an excellent, modern and computer controlled Philips CM3OT TEM/STEM system with a proven high resolution crystal lattice imaging performance. Secondly, the whole column, and not just the region around the sample, has been redesigned for the ECELL functionality. Thirdly, a custom set of polepieces incorporating radial holes for the first stage of differential pumping, but with acceptably low astig matism and no measurable deleterious effect on imaging, has been designed for the instrument.

The basic geometry is a four-aperture system, in pairs above and below the sample (Figs. 2 and 3), but the apertures are now mounted inside the bores of the objective lens polepieces rather than between them as in previous designs. This approach allows unrestricted use of regular sample holders in a relatively narrow gap lens (S = 9 mm) with much lower aberration coefficients ($C_s = C_c = 2$ mm) than have been possible with previous ECELL designs. Regular microscope apertures are mounted in bushes in each polepiece, with 200 lam holes in the inner positions and 400 tm ones outside them. The apertures have direct metal to metal seals around the rim. The aperture openings are large compared to the $\pm$ 20 pm maximum centering error for each aperture (and somewhat less with optical selection). In the manufacture of electron lenses the priority is to minimize misalignment. The absolute sizes are relatively unimportant and it was necessary to measure each bore optically and to construct custom bushes for each position. A measure of self centering was built into the design with conventional elastomer pressure seals. With this approach, standard workshop machining practices were adequate.

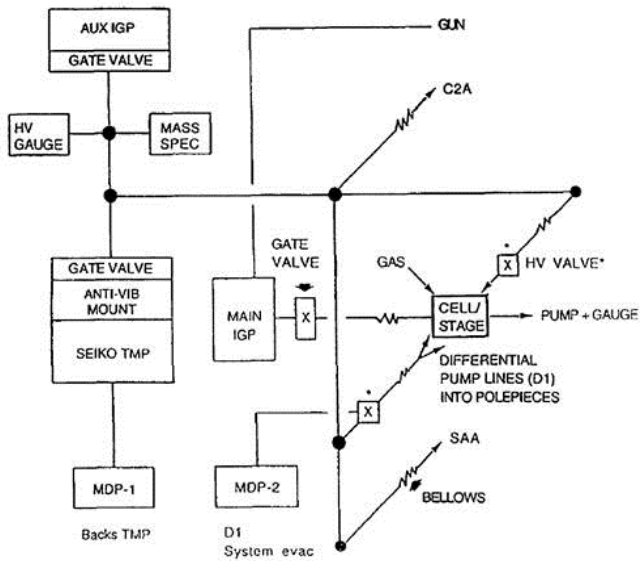

Fig. 2. Schematic of the pumping arrangement on the ECELL-HREM.

The controlled environment ECELL volume is the regular sample chamber of the microscope. It is separated from the rest of the column by the apertures in each polepiece and by the addition of a gate valve, which is normally kept closed, in the line to the regular ion-getter pump (IGP) at the rear of the column. The additional valves are pneumatically operated and interlocked to the main microscope control system. A powerful (340L/S) Seiko turbomolecular pump (TMP) and an auxiliary ion pump have been added to the left side of the microscope [12]. With magnetic TMP bearings, and an additional antivibration mount, no effect on the perfor mance of the microscope has been detected down to 0.2 nm. Both the TMP and the floor mounted molecular drag and diaphragm pump combination (MDP) used to back it are clean (oil-free) systems.

Pumping ports have been added to the column for pumping the sample area of the ECELL, through the side of each polepiece for the first stage of differential pumping between the apertures sealed into each bore, and into the modified column liner tubes for a second stage of pumping at the levels of the condenser lens (C2) and selected area (SA) apertures. We have retained the existing apertures used to separate the column from the electron gun, which is pumped by the original IGP, and from the TEM observation and film camera chamber (with its own oil diffusion pump). The TMP can be connected to pump the main part of the microscope column, including the liner tubes or to evaluate the sample chamber to $5 \times 10^{-8}$ mbar. For ECELL operations the TMP is connected only to maintain a low pressure in the upper and lower linertubes. A second **MDP** system is used to pump inside the polepieces and a diaphragm pump can pump the ECELL gas. Some gas is lost through the quite large apertures in the ECELL and the sample always sees dynamic conditions with a positive flow rate of gas. A separate manifold is used to control the gas supply to the cell (Fig. 4). The construction throughout is stainless steel, and compatible with the UHV and gas pressure conditions. Operating pressures are measured using multiple pirani gauges calibrated for the appropriate gases and disposed around the system to ensure adequate engineering diagnostics. A mass spectrometer is available for inlet gas analysis. Because of the small amounts of solid reactant in the microscope sample, measurement of reaction products are done on larger samples in a microreactor operating under similar conditions and used for microstructural correlation. Electronic image shift and drift compensation help to stabilize high resolution images for data recording on film or with real-time digitally processed video; and minimally invasive electron beam techniques are used throughout.

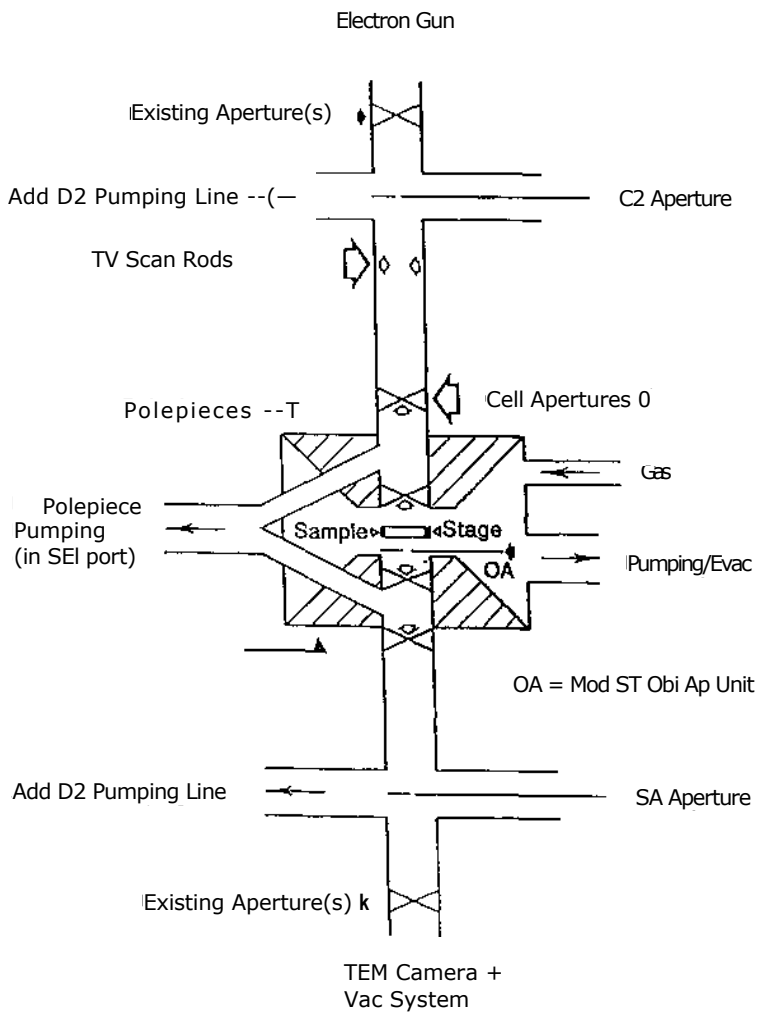

Fig. 3. Schematic of the basic geometry of the aperture system in our design.

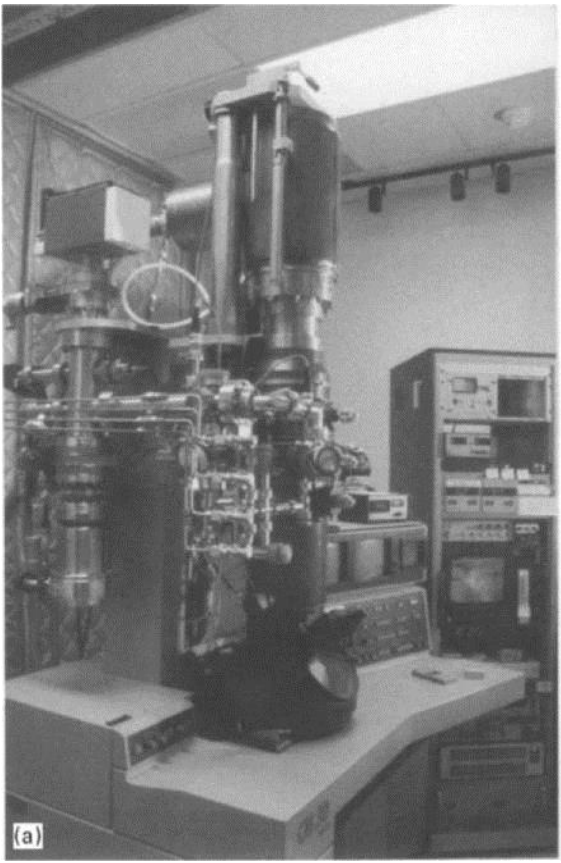
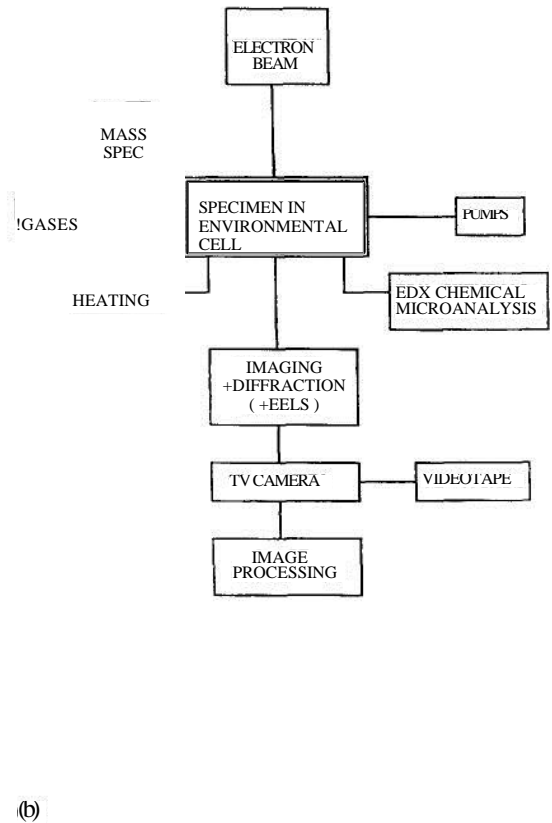

Fig. 4. CM30 ECELL-HRTEM/STEM(EHREM) system constructed at the DuPont Experimental Station. The main turbomolecular pumping system, evacuation and gas handling lines are on the left side of the microscope. Additional pumps, gas supply closets and some electronics are placed behind the acoustic baffle across the middle of the room. (b) Schematic depicting the diverse components of the instrument.

## 3. Results

The alignment and excellent high resolution TEM performance of the microscope have been maintained with the ECELL facilities installed and in operation with sample temperatures above 700°C and with the modest (mbar) amounts of gas flowing through the ECELL. The pressure is limited by scattering in gas environment around the sample and not by limits on the gas load in the system. The tolerance is higher for lighter hydrogen gas than for the heavier nitrogen, air or hydrocarbons. Multi-directional resolution of atomic lattice planes with spacings in the range 0.34-0.2 nm is routine (Fig. 5). Fig. 5a shows an EHREM image with the sample held at 500°C in 0.4 mbar of flowing nitrogen gas. The multi-directional atomic lattice planes of the graphitic carbon with a spacing of 0.34 nm and some gold islands showing 0.23 nm (1 1 1) lattice spacings are clearly resolved under these dynamic conditions. Fig. 5b shows that at 200 kV the ECELL apertures in EHREM restrict the higher angle of diffraction but useful data can be recorded and the high resolution images are not diffraction limited, as evidenced by a well developed island of gold with the (1 1 1) lattice planes resolved in a model gold on amorphous carbon and annealed in situ in a nitrogen atmosphere in the EHREM.

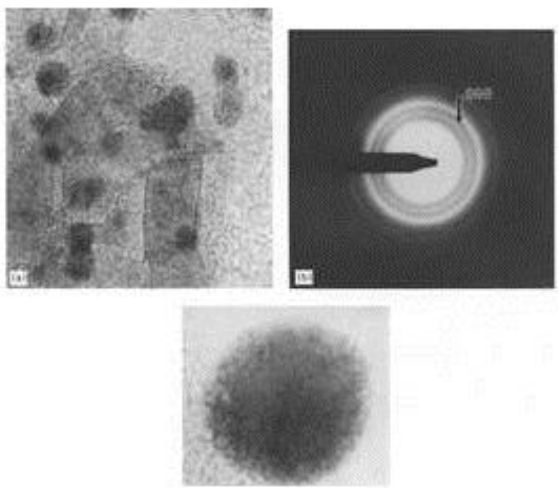

Fig. 5. (a) High resolution TEM image obtained under ECELL operating conditons with the sample held at 500 C in 0.4 mbar of flowing nitrogen gas. The multidirectional atomic lattice planes of the graphitic carbon with a spacing of 0.34 nm are clearly resolved and some gold islands also show the 0.23 nm (1 1 1) lattice spacing. (b) At 200 kV, the ECELL apertures restrict the higher angles of diffraction but useful data can be recorded and the high resolution images are not diffraction limited. (c) A well developed island of gold with the (1 I 1) lattice planes resolved in a model sample of gold on amorphous carbon and annealed at 500 C in the ECELL with nitrogen.

The relatively large apertures in the cell simplify alignment but the main goal is to provide useful angles of diffraction in TEM patterns, and for convergent beam diffraction pattern (CBDP) analysis with a STEM probe. At 200 kV, the patterns extend out to a real space better than 0.07 nm (15 nm and better than 20 nm $^{-1}$ at 300 kV) and do not limit lattice resolution. The regular, smaller, objec-tive apertures can be used inside the EC ELL for diffraction contrast experiments.

## 3.1. Applications to catalysis

The very nature of catalysis with the complex combination of temperature and gas environment presents a formidable challenge in catalytic studies.

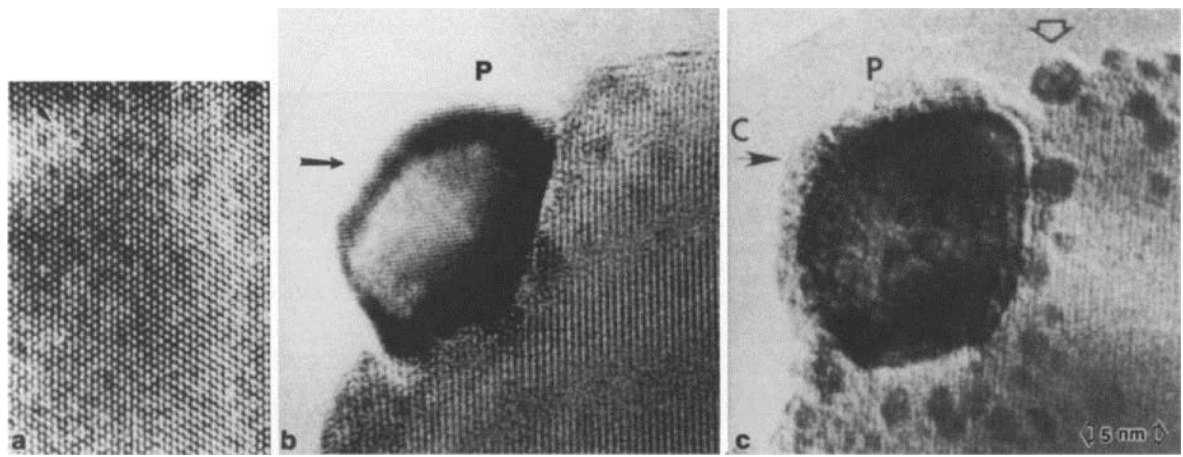

Fig. 6. (a) EHREM of finely dispersed Pt/TiO₂ prepared by impregnation of TiO₂ with Pt solution. (b) In situ dynamic catalyst activation in hydrogen imaged at 300°C. The 0.23 nm (1 1 1) lattice spacings are clearly resolved in the Pt metal particle (P). (c) The same particle of Pt (P) imaged at 450°C, also in H₂. SMSI deactivation with growth of a Ti-rich overlayer (coating close to TiO₂ composition) indicated at C, and the development of nm-scale single crystal clusters of Pt metal (which show no coating as they emerge) (arrowed). All the observations were made in the CM30 ECELL (EHREM) under normal operating conditions of gas and temperature (and representative of the ex situ chemical reaction studies).

EHREM under careful conditions provides a powerful method to study the dynamics of complex real catalysts in powder forms under such extreme conditions.

Strong metal support interactions (SMSI) in a reacting metal particle catalyst can lead to catalyst deactivation. Such phenomena can usefully be examined in EHREM on the atomic scale. An EHREM study of sintering of Pt/rutile titania prepared by impregnation of TiO₂ with Pt solution is shown in Fig. 6. Fig. 6a shows EHREM of finely dispersed Pt/TiO₂ at room temperature (RT). In situ dynamic catalyst activation in hydrogen at 300°C is shown in Fig. 6b. The 0.23 nm lattice spacings are clearly resolved in the Pt metal particle (P). The same particle of Pt (P) (which is partially faceted), is observed under dynamic conditions in hydrogen at — 450°C (Fig. 6c). SMSI deactivation with a growth of an amorphous Ti-rich overlayer or coating, with composition close to slightly anion deficient TiO₂ (as determined by windowless energy dispersive spectroscopy (EDX) and calibrated with standards) is observed (indicated at C), along with the development of nm-scale near spherical single crystal clusters of Pt metal (arrowed). The hydrogen gas is a key contributor to this process.

The in situ results indicate that when the reduction temperature is increased sizes of Pt-particles and the extent of SMSI change. Significantly, the emergence of the nanocrystals imaged during their formation shows that during the early development they are not encapsulated by titania overlayers. Additionally, the Pt particles are not in an epitaxial relationship with the substrate. The results have implications in understanding Pt-dispersion, the role of particle size in SMSI and fraction of catalyst metal available for the gas chemisorption and catalysis. A range of conditions and the dynamic rearrangement of the microstructure can be followed in each in situ experiment. Care was taken to avoid any effects of the electron beam. None were detected and we conclude that this is essentially a noninvasive experiment with respect to beam or vacuum.

### 3.2. EHREM of complex metal oxide catalysts: vanadium—phosphorus—oxygen systems

The selective catalyzation of n-butane to maleic anhydride over vanadium phosphorus oxides is a very important commercial process [13-21], which is an intermediate process to make furans.

The active phase in this selective oxidation is identified as the vanadyl pyrophosphate $(VO)_2P_2O_7$, hereafter referred to as VPO [14]. For good performance, it is necessary that VPO be prepared from vanadyl hydrogen phosphate hydrate precursor, $VOHPO_4 \cdot 1\ H_2O$ (hereafter referred to as VHPO). Transformation of the precursor VHPO to VPO is of considerable importance in the synthesis of the catalyst and has been the subject of a number of studies [14-16]. On an atomic scale, there clearly exists a strong structural association between the two phases, as elucidated by X-ray diffraction studies [17].

The plate-like ("rosette") morphology of VHPO is preserved after the transformation to VPO which has led to suggestions that the transformation is topotactic. However, there is no general consensus on this point based either on energetics, or symmetry. Furthermore, although the precursors and the resulting catalysts have been studied extensively, these indirect studies are primarily post-reaction examinations of the static samples. This is not the same as probing dynamic reacting samples directly under gaseous environments at elevated temperatures. Therefore, uncertainty exists as to the active nature of the transformation. Understanding of the dynamic structural transformation is further hampered because of the lack of microstructural studies of the precursor samples which contain water. The samples are extremely beam sensitive and high-resolution microstructural analyses are difficult to achieve experimentally. A better fundamental understanding of the precursor transformation and temperature is necessary for the development of active catalysts with optimal performance.

We have carried out direct studies of the dynamic transformation of the precursor VHPO to active VPO catalysts, using EHREM under controlled environments. Our observations provide direct evidence about the nature of the dynamic transformation and the associated temperature regimes critical to the formation of active VPO catalysts.

The structure of VHPO [17, 18] consists of vanadyl hydrogen phosphate layers, stacked along the b-axis, which are held together by interlayer hydrogen bonding (Fig. 7a). In this paper, we use a crystallographic notation different from that given in the above references. It is orthorhombic with $a = 7.416$ A, $b = 5.689$ A and $c = 9.592$ A. The layers contain face-sharing $VO_6$ octahedra interconnected by hydrogen phosphate tetrahedra. One of the face-sharing oxygens comes from the water molecule $(H_2O)$ bridging the two vanadium atoms and is located trans to the vanadyl (V-O) oxygen atoms. The structure of VPO [19, 20], related to the precursor VHPO, is shown in Fig. 7b. It is generally agreed that the structure is orthorhombic with $a = 16.694$ A, $b = 7.735$ A and $c = 9.588$ A. Pairs of edge-sharing $VO_6$ octahedra are connected along the b-axis to form double chains of $VO6$ octahedra sharing opposite oxygen corners. Pyrophosphate groups link the double chains into a three dimensional network.

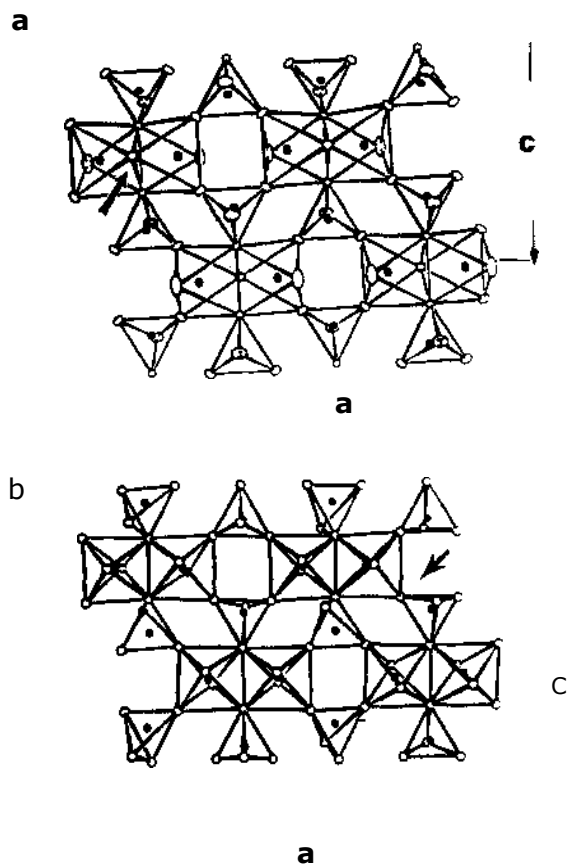

Fig. 7. (a) Structure of the $VOHPO_4 \cdot H_2Oi$ precursor viewed down the $b$ axis. V and P atoms shown as filled circles. One of the bridging $H_2O$ oxygen atoms is marked by an arrow. (b) Structure of $(VO)_2P_2O$, viewed down the $b$ axis. V and P atoms are shown as filled circles. The unit cell is indicated in the (0 1 0) projection and the [2 0 1] direction is arrowed.

Single crystals of VHPO used in the EHREM experiments were prepared from an aqueous media, details of which are given elsewhere [21]. Briefly, they were grown hydrothermally by heating a mixture of 1.05 g of $VO_2$, 0.28 g of $V_2O_3$ and 8 ml of 3 M $H_3PO_4$ at 500°C and slowly cooling the reaction. Phase transformation studies of the single crystals were carried out directly in the EHREM in 2 mbar of flowing nitrogen gas from room temperature (RT) to — 500°C. The in situ dynamic studies are confirmed by blank experiments (without the beam) on samples, with the beam switched on for a few seconds only to record the images of the final state of the material.

The VHPO crystals contain water which presents extreme difficulty in dynamic-EHREM imaging. However, by using controlled gas environments,

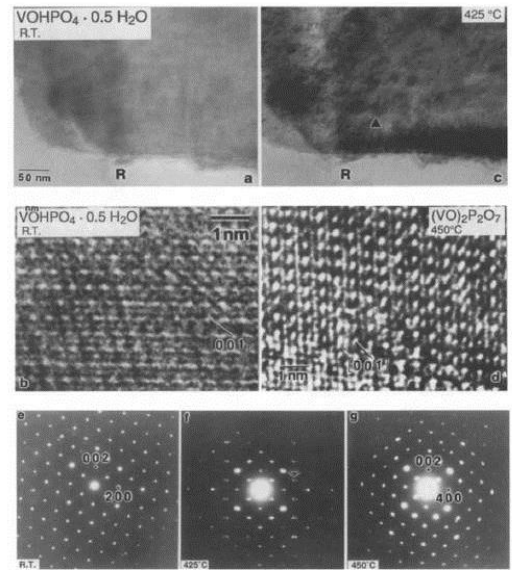

Fig. 8. (a)—(d): In situ transformation of $VOHPO_4 .1\}1_20$ (VHPO) to $(VO)_2P_2O_7$ (VPO), (recorded from the same sample region around R): (a) fresh VHPO containing water, at RT, (b) EHREM structure image of VHPO showing a well-ordered precursor, (c) sample at — 425°C showing decomposition of (0 1 0) VHPO into (0 1 0) VPO microcrystals, and (d) EHREM image at — 450°C elucidating the formation of crystalline VPO. (e)—(g): Dynamic electron diffraction (ED) data of (a), (c), and (d) are shown in (e), (0 and (g), respectively (all are at the same magnification; lattice parameters are given in the text): (e) fresh VHPO containing water, in (0 1 0), (1) intermediate mixture of VHPO and VPO, and (g) final product, VPO, in (0 1 0) orientation. Changes in lattice spacings are evident.

suitably reduced electron gun bias settings, small condenser apertures (30-50 um) and defocused illumination, we have obtained dynamic high resolution images of the reacting samples. Water in the samples does not add to the stability of the catalyst system for imaging but the presence of the gas environment in the ECELL coupled with these imaging conditions minimizes the beam damage. Real-time video recordings were obtained to determine the nature of intermediate species.

The direct EHREM studies show that the structural transformation begins at — 400°C, and a simultaneous existence of the precursor and pyrophosphate phases, both quite crystalline, at — 425 'C. At 450°C, most of the conversion to VPO has taken place. The transformation was completed within an hour. These atomic-scale studies reveal no amorphous phases, or transient defect structures during the transformation, and show that the atomic periodicity is maintained throughout. No other phases have so far been identified in the transformation. These data are summarized in Fig. 8 for a VHPO sample before and during the reaction, recorded from the same sample regions marked R in the figure: (a) fresh VHPO (larger area in the diffraction contrast image) at RT in (0 1 0) crystal orientation, and (b) its HREM structure image indicating a well ordered precursor; (c) the sample at 425°C shows decomposition into (0 1 0) V PO microcrystals in (0 1 0) plane of the host VH PO lattice (hence the low image contrast) and (d) the HREM structure image of the sample at

450°C elucidating the presence of VPO crystals. Samples with (0 0 1) orientation showed dynamic formation of cracks and irregularly shaped VPO microcrystals in the vicinity of the cracks.

Dynamic electron diffraction (ED) data corresponding to the images of VHPO, intermediate species and the final transformation products in Fig. 8a, Fig. 8c and Fig. 8d, are shown in Fig. 8e, Fig. 8f and Fig. 8g, respectively. The EHREM studies show that the microcrystals are well ordered VPO in (0 1 0) orientation which is the catalytically selective basal plane. The transformation is accomplished by the disintegration of VHPO by the loss of water and OH groups into VPO microcrystals. The results provide direct evidence for a topotactic transformation mechanism. The in situ studies are therefore important in understanding reaction mechanisms, in selecting appropriate temperature regimes and in the development of selective catalysts.

We have further carried out catalytic reaction of active VPO in 20% n-butane balanced with He in the EHREM in a gas pressure of 1 mbar. A low voltage SEM micrograph of the VPO catalyst, a schematic of crystal faces in VPO and the relationship between various planes in the VPO lattice are shown in Fig. 9a—Fig. 9c, respectively. EHREM of calcined and activated VPO catalyst in dynamic butane/He reduction are summarized in Fig. 10. The structure image in EHREM of VPO in n-butane/He at RT is shown in the example in Fig. 10a with the corresponding ED pattern shown in Fig. 10b. The surface structural development in butane reduction, followed by diffusion into the bulk, shows the formation of symmetry-related defects (platelets, arrowed) along <2 0 1>. Fig. 10c and Fig. 10d show the defects with the corresponding diffraction pattern with diffuse streaks and Fig. 10e shows one set of defects (marked at D) in high resolution. Under reducing conditions, lattice oxygen loss leads to the formation of coplanar anion vacancies between the vanadyl octahedra and phosphate tetrahedra and the extended defects are introduced along <2 0 1> by glide shear (Fig. 100 to accommodate the misfit strains between the reduced surface layers containing the anion vacancies and the underlying bulk VPO [21]. The novel glide shear mechanism introduces the defects during the reduction of the catalyst but preserves the catalytically active Lewis acid sites required for butane activation and dehydrogenation.

Using EHREM it has been possible to correlate crystal glide shear defects with catalytic activity for the first time. Our earlier in situ EM studies correlated with reaction chemistry [22, 23, 1] have shown that crystallographic shear plane defects (CS) produced by the well known CS mechanism, which eliminate supersaturation of anion vacancies in reducing oxides by shear and lattice collapse, are secondary to catalysis. That is, CS planes are consequences of oxide reactions and not the primary origins of catalytic activity. The catalytically effective novel glide shear mechanism has important implications in understanding how oxide catalysts function. The local oxidation state in the defective regions is estimated as — 3.7-3.8 by measuring the number of defects per unit cell. The defect regions are distributed in VPO which has an oxidation state of 4 [21]. The result is a disordered VPO with defective and unreduced regions differing only in local symmetry. Since disordered glide defects do not lead to a change in the overall VPO structure and tend to oxidize upon exposure to air, these microstructural changes are not readily resolved in X-ray diffraction and in situ electron microscopy and electron diffraction are critical. Dynamic partial oxidation studies of VPO in butane/air environments in the EHREM have also shown the presence of these defects, and no other phases have so far been identified. This suggests the presence of a $V^4 + /V^3$ redox couple on the VPO catalyst surface in the alkane oxidation reaction.

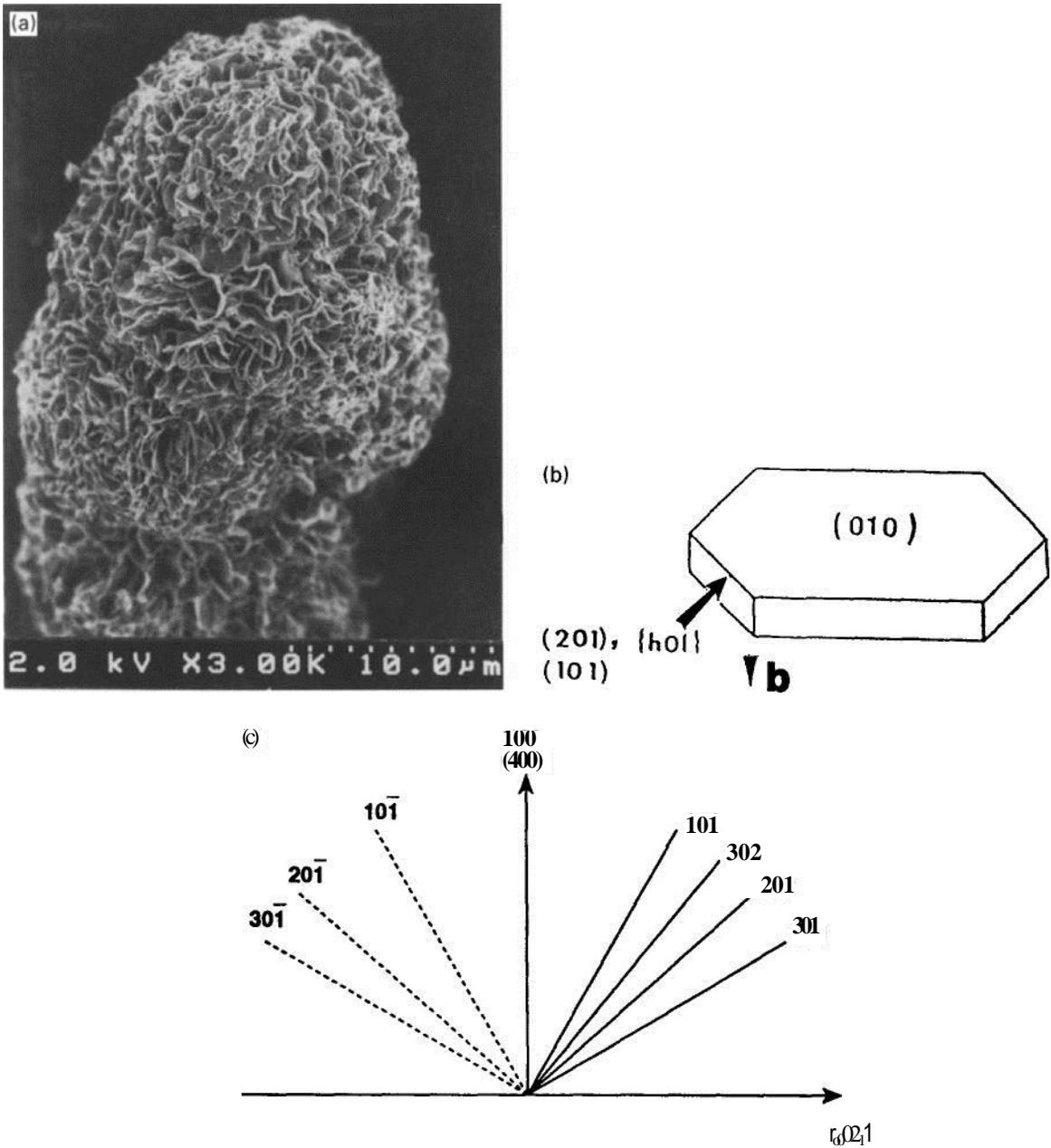

Fig. 9. (a) Low voltage SEM (LVSEM) image of rosette-shaped VPO catalyst, (b) schematic of VPO crystal faces; and (c) relationship between {hkl} in VPO lattice.

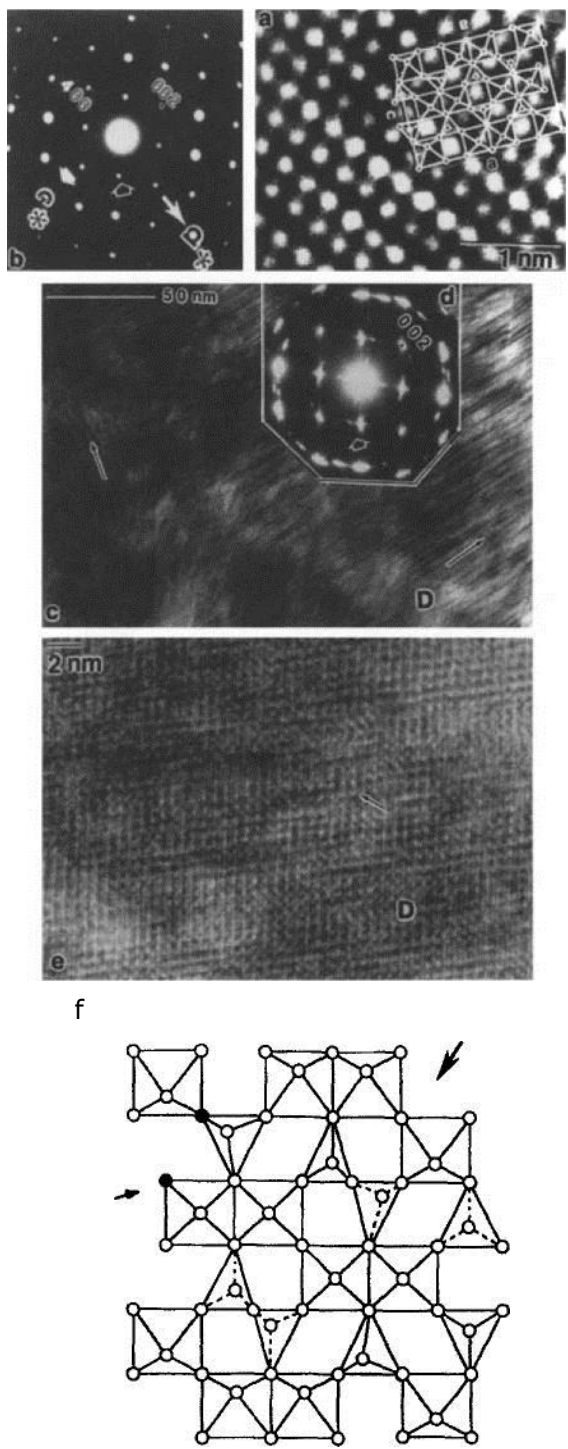

## 4. Conclusions

**A** new in situ environmental cell-HREM (EHREM) has been developed which preserves atomic-scale imaging, diffraction and analytical capabilities under gas environments and at elevated temperatures. Applications to chemical science and catalysis have revealed that in commercially important vanadyl pyrophosphate (VPO) catalysts structural modifications occur by a novel glide shear defect mechanism which maintains catalytically important active Lewis acid sites at the VPO surface. EHREM of transformation of vanadium hydrogen phosphate hydrate precursor to the active VPO catalyst has shown that the transformation is topotactic.

## Acknowledgements

It is a pleasure to thank L.G. Hanna for technical assistance in the assembly and operation of the instrument, D. Sokola for assembly of the external pumping systems, Emile Asselbergs and Philips Electron Optics NV for expert assistance,

Fig. 10. (a) EHREM image of (0 1 0) VPO in n-butane at RT with (b) ED. <2 0 1> reflection is arrowed. (c) In situ EHREM reduction of VPO in n-butane/helium at — 400°C, diffraction contrast showing two sets of defects (arrowed) along <2 0 1> in sample reduced for several hours in situ, with the corresponding electron diffraction (inset, in (d)) which shows diffuse streaks along <2 0 1> (arrowed); (e) EHREM image of one set of the defects (at D) shown in (b). (f) model of glide shear for the extended defects observed in EHREM. Vacant sites are shown by filled circles.

F. Gooding for aperture calibrations, and C.C. Torardi and K. Kourtakis for helpful discussions on VP0 catalysts.